\documentclass[a4paper,11pt]{article}
\usepackage{pos}
\usepackage{cleveref}
\usepackage{natbib}

\newcommand{\lln}{\overline{\ln}}
\newcommand{\nn}{\nonumber}

\title{External-leg corrections as an origin of large logarithms}

\author[a]{Henning Bahl}
\author*[b]{Johannes Braathen}
\author[b,c]{Georg Weiglein}

\affiliation[a]{University of Chicago, Department of Physics, 5720 South Ellis Avenue, Chicago, IL~60637~USA}
\affiliation[b]{Deutsches Elektronen-Synchrotron DESY, Notkestr. 85, 22607 Hamburg, Germany}
\affiliation[c]{II. Institut für Theoretische Physik, Universität Hamburg, Luruper Chaussee 149, 22761 Hamburg, Germany}

\emailAdd{hbahl@uchicago.edu}
\emailAdd{johannes.braathen@desy.de}
\emailAdd{georg.weiglein@desy.de}

\abstract{
Obtaining precise theoretical predictions for both production and decay processes of heavy new particles is of great importance to constrain the allowed parameter spaces of Beyond-the-Standard-Model (BSM) theories, and to properly assess the sensitivity for discoveries and for discriminating between different possible BSM scenarios. 
In this context, it is well known that large logarithmic corrections can appear in the presence of widely separated mass scales. We point out the existence of a new type of possible large, Sudakov-like, logarithms in external-leg corrections of heavy scalars. To the difference of usual Sudakov logarithms, these can furthermore potentially be enhanced by large trilinear couplings. 
Such large logarithms are associated with infrared singularities and we review several techniques to address these at one loop. 
In addition to this discussion, we also present the derivation of the two-loop corrections containing this type of large logarithms, pointing out in this context the importance of adopting an on-shell renormalisation scheme. Finally, we illustrate our calculations and examine the possible magnitude of these corrections for a simple scalar toy model as well as for decay processes involving heavy stop quarks in the Minimal Supersymmetric Standard Model and a heavy Higgs boson in the Next-to-Two-Higgs-Doublet Model.
}

\FullConference{%
  Loops and Legs in Quantum Field Theory - LL2022,\\
  25-30 April, 2022\\
  Ettal, Germany
}

\begin{document}

\begin{flushright}
DESY-22-114
\end{flushright}

\maketitle

\section{Introduction}
The discovery of a 125-GeV Higgs boson at the CERN LHC has completed the particle spectrum of the Standard Model (SM) and confirmed the role of the scalar sector in the realisation of the electroweak phase transition. However, numerous arguments continue to signal the need for Beyond-the-Standard-Model (BSM) physics, both from the experimental side (e.g.\ the baryon asymmetry of the Universe, or the existence of dark matter) or from theory (e.g.\ the gauge hierarchy problem). Many of the BSM theories, devised to address deficiencies of the SM, contain extended scalar sectors -- for instance scalar partners in supersymmetric (SUSY) models, or additional Higgs bosons in both SUSY or bottom-up extensions of the SM. 

In order to correctly assess the discovery sensitivities of BSM scalar and to determine the viable parameter space of BSM models in a reliable way, precise theoretical predictions for the production and decay processes of the new scalars are needed. Meanwhile, the current lack of experimental findings tends to favour\footnote{Another possibility would be that BSM states are light, but have only very small couplings to the SM sector. However, we will not consider this option in the present work.} heavier masses for the BSM states.  

Calculations in Quantum Field Theory are known to be plagued by large logarithmic contributions when widely separated mass scale are present in the considered scenario, leading to a loss of accuracy of the computation. A first example of such terms are logarithms of the ratio of heavy and light mass scales that occur in the calculation of observables at low scales and that can be resummed using Effective-Field-Theory (EFT) techniques -- this is e.g.\ the case for Higgs mass calculations in SUSY scenarios with heavy scalar top (stop) quarks, see the review~\cite{Slavich:2020zjv}. Another type of contributions are Sudakov logarithms, which can appear both in QCD~\cite{Braaten:1980yqDrees:1990dq} and electroweak (see e.g.\ Refs.~\cite{Domingo:2018uimDomingo:2019vit}) computations and can be treated by exponentiation or using techniques from Soft-Colinear Effective Theory (SCET)~\cite{SCET} -- see also Refs.~\cite{SCETwBSM} for some (model-specific) applications of SCET techniques to the decay of BSM particles into SM ones.  
In Ref.~\cite{Bahl:2021rts}, we pointed out the existence of a new type of large Sudakov-like logarithmic contributions, arising in external-leg corrections of heavy scalars and that can be enhanced by sizeable trilinear couplings. We carried out a detailed analysis of the origin and possible magnitude of these terms, which we summarise in these proceedings. 
\vspace{-.3cm}

\section{Large logarithms in external-leg corrections}
To illustrate in a clear manner our discussion of large logarithmic contributions in external-leg corrections, we consider first a toy model of three real scalars, $\phi_i$ ($i=1,2,3$), and one Dirac fermion $\chi$. The model is endowed with a global, unbroken, $\mathbb{Z}_2$ symmetry under which $\phi_1\to-\phi_1$, $\phi_2\to-\phi_2$, $\phi_3\to\phi_3$, and $\chi\to\chi$. In the following, we consider hierarchical mass scenarios, in which $\phi_1$ is light (or massless) while $\phi_2$ and $\phi_3$ are heavy and at approximately the same mass ($m_2\sim m_3$). 
Due to the $\mathbb{Z}_2$ symmetry, the possible couplings of the model are constrained. In particular, only $\phi_3$ can couple to fermions. Moreover, for the discussion of large logarithms, trilinear scalar couplings will play a central role, and among them the most important is the light-heavy-heavy coupling between $\phi_1$, $\phi_2$, $\phi_3$, which we denote $A_{123}$. To summarise, the interaction terms that will play an important role in the following are
\begin{align}
 \mathcal{L}_\text{int.}\supset -A_{123}\phi_1\phi_2\phi_3+y_3\bar\chi\chi\phi_3\,. 
\end{align}

As a prototype of decays of a scalar into two fermions, or of a fermion into a scalar and a fermion, in realistic models, we investigate the decay $\phi_3\to\bar\chi\chi$. At tree level, the calculation of the decay width for this process is straightforward and yields $\Gamma^{(0)}(\phi_3\to\bar\chi\chi)=m_3y_3^2/(8\pi)(1-4m_\chi^2/m_3^2)^{3/2}$, where $m_3$ and $m_\chi$ denote the masses of $\phi_3$ and of $\chi$. From the one-loop order, one must also include external scalar- and fermion-leg as well as vertex corrections, however restricting our attention exclusively to corrections involving trilinear couplings, only external-leg corrections of scalars need to be considered. These are included via LSZ factors and can be written (up to two loops) as
\begin{align}
\label{EQ:gen_phi3decay}
    \hat\Gamma(\phi_3\to\chi\bar\chi)=\Gamma^{(0)}(\phi_3\to\chi\bar\chi)\bigg[&1+\Delta\hat\Gamma^{(1)}+\Delta\hat\Gamma^{(2)}\bigg]\\
    =\Gamma^{(0)}(\phi_3\to\chi\bar\chi)\bigg[&1-\mathfrak{Re}\hat\Sigma_{33}^{(1)\prime}(m_3^2)-\mathfrak{Re}\hat\Sigma_{33}^{(2)\prime}(m_3^2)+\big(\mathfrak{Re}\hat\Sigma_{33}^{(1)\prime}(m_3^2)\big)^2\nn\\
            -&\frac12\big(\mathfrak{Im}\hat\Sigma_{33}^{(1)\prime}(m_3^2)\big)^2+\mathfrak{Im}\hat\Sigma_{33}^{(1)}(m_3^2)\cdot\mathfrak{Im}\hat\Sigma_{33}^{(1)\prime\prime}(m_3^2)+\mathcal{O}(\text{3-loops})\bigg]\,,\nn
\end{align}
where $\hat\Sigma_{33}^{(1)}(m_3^2)$ ($\hat\Sigma_{33}^{(2)}(m_3^2)$) denotes the renormalised one-loop (two-loop) self-energy of $\phi_3$ evaluated at external momentum equal to $p^2=m_3^2$ and the prime denotes the derivative with respect with respect to $p^2$.  
At one-loop order, the term involving $A_{123}$, namely
\begin{align}
 \Delta\hat\Gamma^{(1)}\supset-\frac{1}{16\pi^2}\mathfrak{Re}\bigg[(A_{123})^2\frac{d}{dp^2}B_0(p^2,m_1^2,m_2^2)\bigg]_{p^2=m_3^2}\,,
\end{align} 
becomes infrared (IR) divergent if $m_1\to 0$ and $m_2\to m_3$ (in this expression, $B_0$ is the usual Passarino-Veltman function~\cite{Passarino:1978jh}). If on the one hand $m_2=m_3$, then the IR divergence is regulated by $m_1$ and the derivative of the $B_0$ function becomes $\frac{d}{dp^2}B_0(p^2,m_1^2,m_3^2)|_{p^2=m_3^2}=[1/2\log (m_3^2/m_1^2)-1+\mathcal{O}(m_1^2/m_3^2)]/m_3^2$. On the other hand, when $m_1=0$, the IR divergence is regulated by the difference of the squared masses of $\phi_2$ and $\phi_3$ and the $B_0$ derivative is $\frac{d}{dp^2}B_0(p^2,0,m_2^2)|_{p^2=m_3^2}=\big[\log\frac{m_3^2}{m_2^2-m_3^2}-1+\mathcal{O}(\frac{m_2^2-m_3^2}{m_3^2})\big]/m_3^2$. A further complication associated with this term is that the apparent suppression by $m_3^{-2}$ is compensated by the prefactor $A_{123}^2\sim m_3^2$.

The appearance of IR divergences in the computation of a decay width is, however, not an unknown phenomenon, and several methods can allow treating these. A first option, inspired by solutions to the Goldstone boson catastrophe (see Refs.~\cite{GBC}), is to resum contributions involving the light scalar $\phi_1$. With this approach, the IR divergence is interpreted as stemming from an inappropriate perturbative expansion -- because in a scenario with a large mass hierarchy, the light scalar mass receives significant radiative corrections, meaning that in turn diagrams with an insertion of a $\phi_1$ self-energy will typically be large. However, the physical interpretation of the resummed decay width, in particular in terms of what physical observable it should be compared to, is not clear. Another approach is to apply the Kinoshita-Lee-Nauenberg theorem~\cite{KLN} and include in the calculation the soft $\phi_1$ radiation process $\phi_2\to\phi_1\bar\chi\chi$ -- thereby interpreting the IR divergence as stemming from a lack of inclusiveness of the computed observable. Nevertheless, the situation can occur that either $m_1$ or $m_2^2-m_3^2$ are large enough for the $\phi_3\to\bar\chi\chi$ and $\phi_2\to\phi_1\bar\chi\chi$ processes to be distinguished experimentally. In such a case, a large and unsupressed -- due to the $(A_{123}/m_3)^2\sim 1$ prefactor -- logarithmic term remains in the one-loop contribution to the decay width. This poses the questions of the size of similar effects in higher-order contributions, and of whether some resummation, for instance using techniques inspired by SCET, would be necessary. 

To answer these questions, we compute the two-loop external-leg contributions of order $\mathcal{O}(A_{123}^4)$. These involve derivatives with respect to external momentum of two-loop self-energy integrals, namely $T_{11234}(m_2^2,m_2^2,m_1^2,m_3^2,m_1^2)$, $T_{11234}(m_1^2,m_1^2,m_2^2,m_2^2,m_3^2)$, and $T_{12345}(m_2^2,m_1^2,m_3^2,m_1^2,m_2^2)$ (using the $T$-integral notations from e.g.\ Ref.~\cite{Weiglein:1993hd}), evaluated at $p^2=m_3^2$. Although for finite external momenta analytical expressions for two-loop self-energy integrals, or their derivatives, are not known for general internal masses, we calculate for the case $m_1^2=\epsilon$ and $m_2^2=m_3^2=m^2$ the leading pieces of these derivatives in powers of $\epsilon/m^2$, employing the systems of differential equations between self-energy integrals as well as analytical expressions from Ref.~\cite{Martin:2003qz} together with results from  Refs.~\cite{Martin:2003itMartin:2005qm}. Combining these results, we obtain for the two-loop corrections to the $\phi_3\to\bar\chi\chi$ decay width in terms of $\overline{\text{MS}}$-renormalised parameters
\begin{align}
\label{EQ:MS_2L}
 \Delta\hat\Gamma^{(2)}\supset&\  \frac{A_{123}^4}{256\pi^4m^4}\bigg[\frac{m^2\lln m^2}{2 \epsilon} - \frac{m \pi (4 + \lln m^2)}{8 \sqrt{\epsilon}}+\frac{17}{9}-\frac{\pi^2}{8}+\frac{1}{8}\ln^2\frac{m^2}{\epsilon}+\frac{1}{6}\lln\epsilon+\frac{1}{12}\lln m^2\nn\\
            &\hspace{2.25cm}+\pi^2\ln 2-\frac32 \zeta(3)\bigg]\,,
\end{align}
where $\lln x\equiv\ln x/Q^2$, $Q$ being the renormalisation scale. In addition to a $\ln\epsilon$ term, this $\overline{\text{MS}}$ result contains a $\ln^2\epsilon$ term as well as dangerous power-enhanced terms of the form $m^2/\epsilon$ and $m/\sqrt{\epsilon}$. 
Other renormalisation schemes will however prove more appropriate, as we will see in the following numerical examples. The finite counterterms\footnote{Note that we choose to retain an $\overline{\text{MS}}$ prescription for the field renormalisation.} for the scalar masses ($\delta^{(1)}m_1^2$ and $\delta^{(1)}m_2^2$) and for the trilinear coupling $A_{123}$ ($\delta^{(1)}A_{123}$) enter via the following subloop-renormalisation contributions
\begin{align}
   \hat\Sigma_{33}^{(2\text{, subloop})}(p^2)=\frac{A_{123}^2}{256\pi^4}\bigg[&\frac{2\delta^{(1)}A_{123}}{A_{123}}B_0(p^2,m_1^2,m_2^2)  + \delta^{(1)}m_1^2\frac{\partial}{\partial m_1^2}B_0(p^2,m_1^2,m_2^2)\nn\\
   & + \delta^{(1)}m_2^2\frac{\partial}{\partial m_2^2}B_0(p^2,m_1^2,m_2^2)\bigg]\,.
\end{align}
We find that the contributions from on-shell (OS) mass counterterms have precisely the form required to cancel the dangerous $1/\epsilon$ and $1/\sqrt{\epsilon}$ terms in the two-loop external-leg corrections stemming from genuine two-loop diagrams -- see \cref{EQ:MS_2L}. Next, for the trilinear coupling, different schemes can be considered: \textit{(i)} first, the simplest possibility would be to keep an $\overline{\text{MS}}$ renormalisation, i.e.\ $\delta^{(1)}A_{123}\big|^\text{finite}=0$; \textit{(ii)} we can adopt a (process-dependent) OS scheme for $A_{123}$ and fix $\delta^{(1)}A_{123}\big|^\text{finite}$ by the requirement of ensuring that the OS-renormalised loop-corrected amplitude for the $\phi_2\to\phi_1\phi_3$ process remains equal to the tree-level amplitude; \textit{(iii)} we can devise a custom renormalisation scheme, which we name ``no-log-sq.'', to cancel the $\ln^2\epsilon$ term in the two-loop external-leg corrections to the $\phi_3\to\bar\chi\chi$ decay width (it should be emphasised that the $\ln^2\epsilon$ does not disappear entirely from the computation, and that it would reappear if one were to extract the value of $A_{123}$ in this scheme from a physical observable). Detailed results for $\Delta\hat\Gamma^{(2)}$ in these different schemes are provided in Ref.~\cite{Bahl:2021rts}. 

\begin{figure}
 \centering
 \includegraphics[width=.9\textwidth]{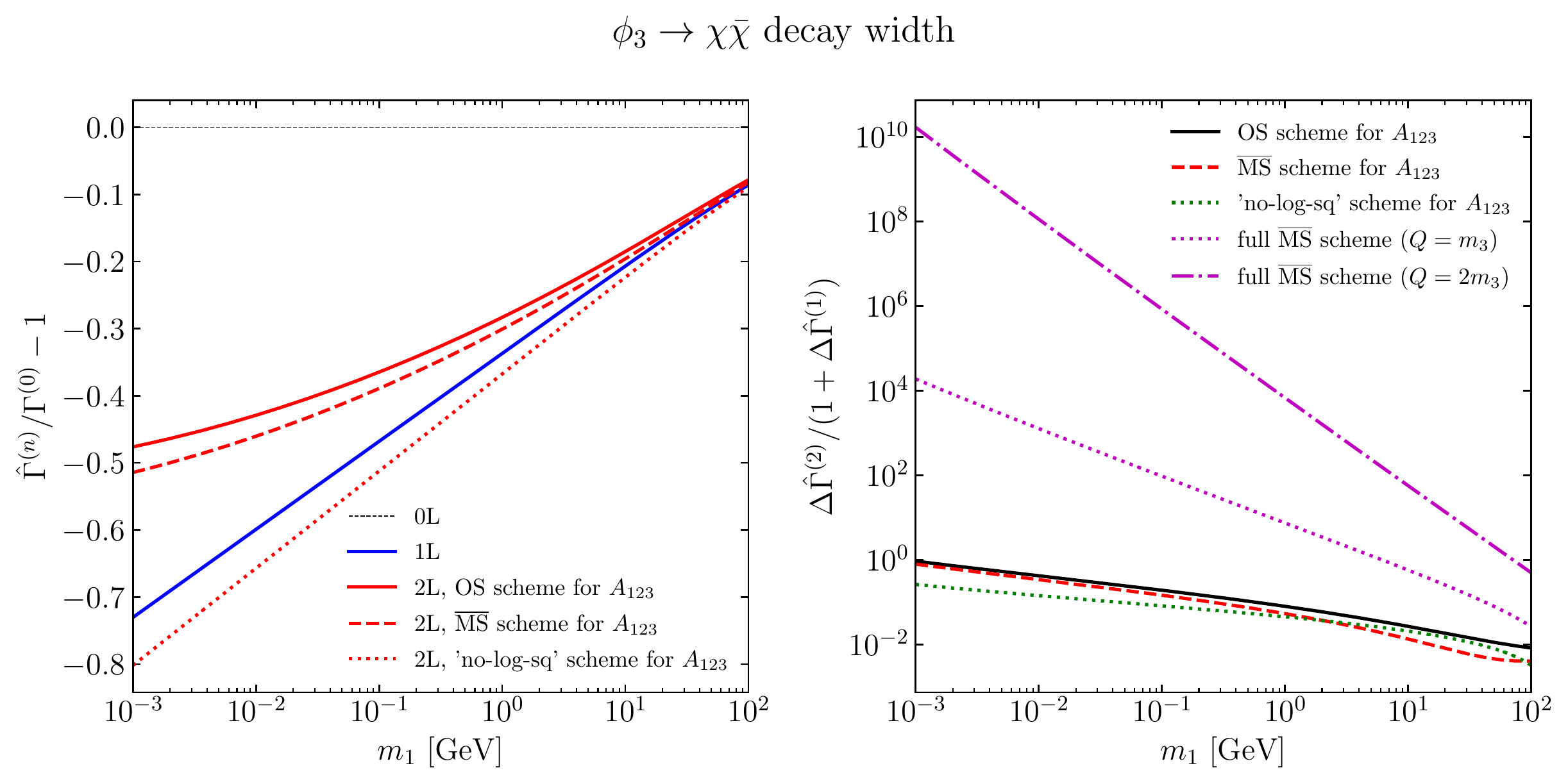}
 \caption{The one- and two-loop external $\phi_3$ leg corrections to the $\phi_3\to\bar\chi\chi$ decay width as a function of $m_1$. \textit{Left}: one- and two-loop results for the decay width, shown relative to the tree-level result; \textit{right}: two-loop corrections shown relative to the one-loop result. We have chosen $m_2 = m_3 = 1 \text{ TeV}, y_3 = 1, A_{123} = 3 \text{ TeV}$, while the other trilinear couplings are set to zero.}
 \label{FIG:ToyModel}
\end{figure}

In the left plot of \cref{FIG:ToyModel}, we present the size of the one- (blue curve) and two-loop (red curves) external-leg corrections to the $\phi_3\to\bar\chi\chi$ decay width, relative to the tree-level prediction, as a function of the light mass $m_1$. We consider a parameter point with $m_2=m_3=1\text{ TeV}$, $y_3=1$, and $A_{123}=3\text{ TeV}$, while other trilinear couplings are set to zero (note that the red curves correspond to different physical points because the renormalisation-scheme interpretation of $A_{123}$ differs between them). The straight line for the one-loop result with the semi-logarithmic scale corresponds exactly to the logarithmic divergence with $\epsilon=m_1^2$. At two loops, the $\overline{\text{MS}}$ and OS results (dashed and solid curves respectively) are similar and exhibit a large impact from the two-loop corrections -- for $m_1=1\text{ MeV}$ they amount to about $-1/3$ of the one-loop effects -- because of the $\ln^2\epsilon$ terms. In the ``no-log-sq.'' scheme, the two-loop corrections are, as expected, significantly smaller (e.g.\ only $7\%$ of the one-loop contributions for $m_1=1\text{ MeV}$) due to the absence of a $\ln^2\epsilon$ piece. In the right plot of \cref{FIG:ToyModel}, we compare the size of the two-loop external-leg corrections to the same $\phi_3\to\bar\chi\chi$ decay width relative to the one-loop result, as a function of $m_1$, for different choices of renormalisation prescriptions for the scalar masses and $A_{123}$. For the black, red, and green curves, we renormalise the scalar masses $m_1$ and $m_2$ in the OS scheme, while for $A_{123}$ we use respectively the OS, $\overline{\text{MS}}$, and ``no-log-sq.'' schemes. For the magenta curves, we renormalise both the scalar masses and the trilinear coupling in the $\overline{\text{MS}}$ scheme, taking two different values for the renormalisation scale $Q$, respectively $Q=m_3$ (dotted curve) and $Q=2m_3$ (dot-dashed curve). As long as the scalar masses are renormalised on-shell, the relative magnitude of the two-loop contributions remains moderate (again, the custom ``no-log-sq.'' scheme leads to the the smallest effects). However, the magenta curves illustrate the breakdown of the full $\overline{\text{MS}}$ calculation due to the $1/\epsilon$ and $1/\sqrt{\epsilon}$ terms: indeed the two-loop contributions are both unphysically large and show a huge dependence on the renormalisation scale. This example illustrates the importance of adopting an OS renormalisation of masses in the decay width calculation (similar observations were made for instance in Refs.~\cite{Nondecoupling} in the context of Higgs mass calculations), and moreover poses the question of the size of external-leg corrections at one and two loops in realistic models, to which we turn now. 
\vspace{-.3cm}

\section{Numerical investigations in the MSSM and N2HDM}\vspace{-.2cm}
\begin{figure}
 \centering
 \includegraphics[width=.9\textwidth]{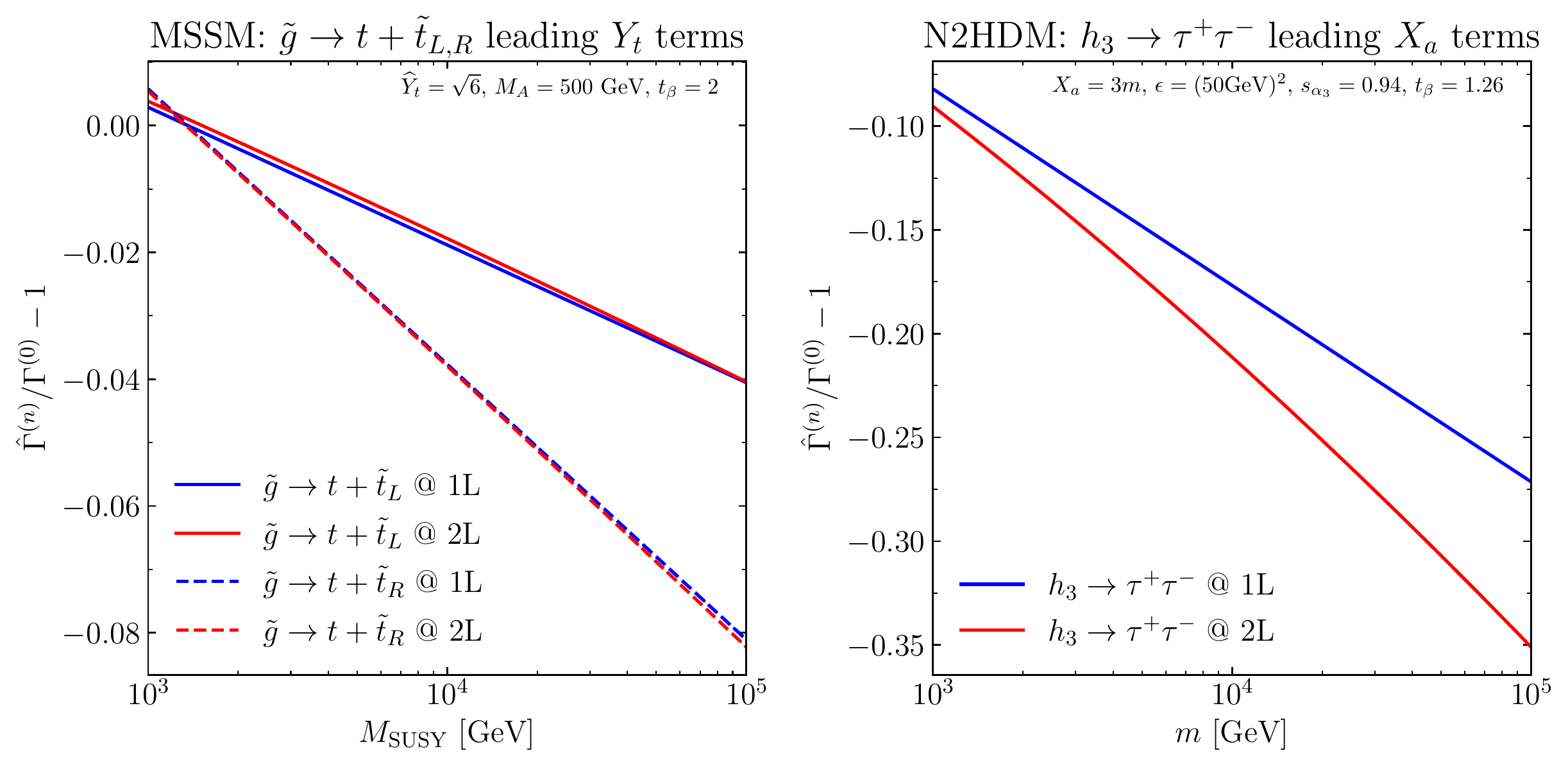}
 \caption{\textit{Left:} Leading $Y_t$ external-leg corrections to the decay of a gluino into a top quark and a left- or right-handed stop quark in the MSSM. \textit{Right:} Leading external-leg contributions in powers of $X_a$ to the $h_3\to\tau^+\tau^-$ decay in the N2HDM. For both plots, one-loop results are shown in blue, while two-loop ones are in red. The choices of parameters for the two models are given in the figures.}
 \label{FIG:MSSM+N2HDM}
\end{figure}

In this section, we review some examples of results for external-leg contributions to realistic decay processes in the Minimal Supersymmetric SM (MSSM) and the Next-to-Two-Higgs-Doublet Model (N2HDM) -- for both models we assume CP conservation. We refer the reader to Ref.~\cite{Bahl:2021rts} for our definitions and conventions for these two models, as well as for our analytical results. 
First, in the MSSM, we investigate the decay of a gluino into a top quark and a stop quark (either left- or right-handed). Specifically, we consider scalar (stops in this case) external-leg contributions involving the coupling $Y_t\equiv A_t+\mu\cot\beta$ -- where $A_t$ is the trilinear stop coupling, $\mu$ is the Higgsino mass parameter and $\cot\beta=1/t_\beta$ with $t_\beta$ the ratio of the vacuum expectation values of the neutral components of the two Higgs doublets. The parameter $Y_t$ enters in couplings between stop quarks and the BSM scalars of the MSSM. Consequently, the mass hierarchy that we study is one where the stop quarks are heavy and degenerate at a scale $M_\text{SUSY}$ that will be varied between 1 and 100 TeV while the BSM scalars play the role of the light (but not massless) scalars, with a fixed mass scale of $M_A=500\text{ GeV}$. We present in the left hand plot of \cref{FIG:MSSM+N2HDM} our results for the one- (blue curves) and two-loop (red curves) corrections to the gluino decay as a function of the heavy mass scale $M_\text{SUSY}$, once again in relative size to the tree-level result and working in an OS scheme for the scalar masses and $Y_t$. The solid (dashed) lines indicate the results for the decay involving a left- (right-) handed stop quark. In addition to $M_A=500\text{ GeV}$, we fix for this figure $Y_t=\sqrt{6}M_\text{SUSY}$ and $t_\beta=2$. We find that the one-loop corrections are moderate, giving rise to deviations of a few percent from the tree-level result, while the two-loop effects are minute. 

Turning next to the N2HDM, we investigate contributions to the decay of the heaviest CP-even scalar $h_3$ into a pair of tau leptons involving the coupling $X_a$. We define $X_a$ in terms of Lagrangian trilinear couplings as $X_a\equiv(a_{1S}-a_{2S})/4$ -- where $a_{1S},\ a_{2S}$ are trilinear couplings between the two doublets and the singlet of the N2HDM, defined as $\mathcal{L}_\text{int.}\supset-a_{1S}|\Phi_1|^2\Phi_S/2-a_{2S}|\Phi_2|^2\Phi_S/2$. We focus on a scenario with a mass hierarchy where the first two CP-even states $h_1,\ h_2$ and the would-be Goldstone bosons are the light states -- for simplicity we fix them at a common scale $\sqrt{\epsilon}=50\text{ GeV}$ -- while the third CP-even as well as the CP-odd and charged Higgs bosons are heavy and degenerate at a scale $m$. We furthermore choose a large trilinear coupling $X_a=3m$, fix $t_\beta=1.26$ and we ensure that the mixing angle $\alpha_3$ is such that the factors of $\cos{\alpha_3}$ associated with each power of the coupling $X_a$ (c.f.\ expressions in Ref.~\cite{Bahl:2021rts}) are not too suppressed while still evading experimental limits -- specifically we take $\sin{\alpha_3}=0.94$. The results that we obtain are shown in the right plot of \cref{FIG:MSSM+N2HDM}, as a function of the heavy mass scale $m$ (varied between 1 and 100 TeV). The blue curve indicates the one-loop result relative to the tree-level one, while the red curve includes also the two-loop effects. For this scenario, we find that the one- and two-loop external-leg corrections can be quite significant, due to the large number of diagrams contributing in the N2HDM -- the one-loop effects become as large as $\sim-27\%$ for $m=100\text{ TeV}$ while the two-loop contributions produce an additional shift of $-8\%$. Although the external-leg corrections are larger in the N2HDM than in the MSSM (due to the larger number of diagrams contributing in the former model) the two-loop corrections remain well smaller than their one-loop counterparts, and the perturbative expansion is clearly well-behaved. In conclusion, for both realistic models considered in this work, a fixed-order computation appears to remain sufficient to obtain a reliable result for the investigated decay processes, and it does not appear mandatory to develop a SCET-like resummation of the logarithmic contributions. 

\vspace{-.3cm}
\section{Summary}
\vspace{-.2cm}
Precise theory predictions are of paramount importance to properly assess BSM discovery sensitivities, and to constrain parameter spaces of BSM models in light of experimental searches for new states. In Ref.~\cite{Bahl:2021rts}, we pointed out the existence of a new type of large Sudakov-like logarithms, appearing in
external-leg corrections of heavy scalars in presence of a mass hierarchy between scalars. In contrast to usual Sudakov logarithms, these can be further enhanced by large trilinear couplings. Working first at one-loop order, we showed how these logarithmic terms are related to singularities in the IR limit, and we reviewed different methods to address the associated divergences. Next, we computed the two-loop external-leg contributions involving large logarithms. In this context, we demonstrated the importance of adopting an OS renormalisation scheme for the scalar masses, in order to avoid unphysical enhancements of the corrections to the decay width. Finally, we illustrated our results by examples of contributions to the decay of a gluino into a top quark and a stop in the MSSM and to the decay of a heavy scalar into tau leptons in the N2HDM -- further examples and scenarios can be found in Ref.~\cite{Bahl:2021rts}. While we find potentially large effects at one loop, the two-loop logarithmic terms remain well smaller than the one-loop contributions, meaning that a SCET resummation does not seem compulsory to obtain reliable predictions.

\vspace{-.3cm}
\section*{Acknowledgements}
\sloppy{We acknowledge support by the Deutsche Forschungsgemeinschaft (DFG, German Research Foundation) under Germany‘s Excellence Strategy -- EXC 2121 ``Quantum Universe'' – 390833306. H.B.\ acknowledges support by the Alexander von Humboldt foundation. }


\end{document}